\documentclass[aps,twocolumn,prd,amssymb,eqsecnum,nofootinbib,floatfix]{revtex4}
\usepackage{epsfig}
\usepackage{amsmath}
\usepackage{amsfonts}
\usepackage{bm}
\usepackage{revsymb}
\usepackage{graphicx,epsfig}
\usepackage{placeins}
\newcommand{\be}{\begin{equation}}
\newcommand{\ee}{\end{equation}}
\newcommand{\bea}{\begin{eqnarray}}
\newcommand{\eea}{\end{eqnarray}}
\newcommand{\bes}{\begin{subequations}}
\newcommand{\ees}{\end{subequations}}
\newcommand{\nn}{\nonumber}

\begin{document}
\interfootnotelinepenalty=10000

\title{Can Gravity Probe B usefully constrain torsion gravity theories?}

%
%
\newcount\hh
\newcount\mm
\mm=\time
\hh=\time
\divide\hh by 60
\divide\mm by 60
\multiply\mm by 60
\mm=-\mm
\advance\mm by \time
\def\hhmm{\number\hh:\ifnum\mm<10{}0\fi\number\mm}


\author{\'Eanna \'E. Flanagan, Eran Rosenthal}
\affiliation{Center for Radiophysics and Space Research, Cornell
  University, Ithaca, New York, 14853}
\date{draft of May 8, 2007; printed \today{} at \hhmm}

\begin{abstract}

In most theories of gravity involving torsion, the source for torsion is
the intrinsic spin of matter.  Since the spins of fermions
are normally randomly oriented in macroscopic bodies, the amount of
torsion generated by macroscopic bodies is normally negligible.
However, in a recent paper, Mao et al. (gr-qc/0608121) point out that there is
a class of theories, including the Hayashi-Shirafuji (1979) theory, in
which the angular momentum of macroscopic spinning bodies generates a
significant amount of torsion.
They further argue that, by the principle of action equals reaction,
one would expect the angular momentum of test bodies to couple to a
background torsion field, and therefore the precession of the Gravity
Probe B gyroscopes should be affected in these theories by the torsion
generated by the Earth.

We show that in fact the principle of action equals reaction does not apply to
these theories, essentially because the torsion is not an independent
dynamical degree of freedom.  We examine in detail a generalization of
the Hayashi-Shirafuji theory suggested by Mao et al. called
Einstein-Hayashi-Shirafuji theory.
There are a variety of different versions of this theory,
depending on the precise form of the coupling to matter chosen for the torsion.
We show that for any coupling to matter that is compatible with
the spin transport equation postulated by Mao et al., the theory
has either ghosts or an ill-posed initial value formulation.
These theoretical problems can be avoided by specializing the parameters
of the theory and in addition choosing the standard minimal coupling
to matter of the torsion tensor.  This yields a
consistent theory, but one in which the action equals reaction
principle is violated, and in which
the angular momentum of
the gyroscopes does not couple to the Earth's torsion field.
Thus, the Einstein-Hayashi-Shirafuji theory does not predict a
detectable torsion signal for Gravity Probe B.  There may be other
torsion theories which do.

\end{abstract}

\maketitle

\section{Introduction and Summary}
\label{sec:intro}

\subsection{Theories of gravity with torsion}
\label{sec:torsiontheories}

General relativity (GR) is in good agreement with all current
experimental data from laboratory tests, the Solar System\footnote{An exception
is the Pioneer anomaly \cite{Nieto:2005kb}, which remains controversial.}
and binary pulsars \cite{lrr-2001-4}.  However there is good
motivation to consider modifications and extensions of GR: low energy
limits of string theory and higher dimensional models usually involve
extra, long range universal forces mediated by scalar fields, and in
addition the observed acceleration of the Universe  may be due to a
modification of GR at large distances. One may hope that
new and highly accurate experiments, such as Gravity Probe B \cite{GPB},  will
enable one  to test for deviations from GR.

One natural framework in which to generalize GR is to allow the
connection $\Gamma^\mu_{\ \nu\lambda}$ to be a nonsymmetric  independent
dynamical variable instead of being determined by the metric.
The covariant derivative of a vector $v^\mu$ is defined in the
usual way as
\be
\nabla_\mu v^\nu = \partial_\mu v^\nu + \Gamma^\nu_{\
  \mu\lambda} v^\lambda.
\ee
If one retains the assumption that the connection is metric
compatible, $\nabla_\mu g_{\nu\lambda}=0$,
then it can be shown that the connection is determined uniquely by the metric
and the torsion tensor
\be
S_{\mu\nu}^{\ \ \,\lambda} \equiv \Gamma^{\lambda}_{\ [\mu \nu]}.
\ee
One obtains
\be
\Gamma^\lambda_{\ \mu\nu} = \left\{ {}^{\,\,\lambda}_{\mu\nu}
\right\} - K_{\mu\nu}^{\ \ \,\lambda}
\label{kdef}
\ee
where the first term is the Levi-Civita connection determined by the
metric and
\be
K_{\mu\nu}^{\ \ \,\lambda}=-S_{\mu\nu}^{\ \ \,\lambda} -S_{\
  \,\nu\mu}^{\lambda} -S_{\ \,\mu\nu}^{\lambda}
\label{kands}
\ee
is called the contorsion tensor.
A spacetime equipped with a metric $g_{\mu\nu}$ and a torsion tensor is called a
Riemann-Cartan spacetime.
The Riemann tensor $R^\mu_{\ \, \nu\lambda\rho}$ of the full
connection (\ref{kdef})
is related to the usual Riemann tensor ${\tilde R}^\mu_{\ \,\nu\lambda\rho}$
of the Levi-Civita connection by
\bea
R^\mu_{\ \, \nu\lambda\rho}  &=& {\tilde R}^\mu_{\ \, \nu\lambda\rho}
+ {\tilde \nabla}_\rho K_{\lambda\nu}^{\ \ \,\mu}
- {\tilde \nabla}_\lambda K_{\rho\nu}^{\ \ \,\mu}
+ K_{\lambda\sigma}^{\ \ \,\mu} K_{\rho\nu}^{\ \ \,\sigma} \nonumber \\
&&- K_{\rho\sigma}^{\ \ \,\mu} K_{\lambda\nu}^{\ \ \,\sigma},
\label{eq:Riemann1}
\eea
where ${\tilde \nabla}_\mu$ is the Levi-Civita derivative operator, and our convention for the Riemann tensor
 is given by Eq. (\ref{barriemann}).
The action for theories of gravity in this framework has the generic form
\begin{equation}
\label{fullaction0}
S[g_{\mu\nu},\Gamma^\lambda_{\
  \mu\nu},\Psi]=S_G[g_{\mu\nu},\Gamma^\lambda_{\
  \mu\nu}] +S_{\rm matter}[g_{\mu\nu},\Gamma^\lambda_{\
  \mu\nu},\Psi ] \,,
\end{equation}
where $S_G$ and $S_{\rm matter}$ are the gravitational and matter
actions and $\Psi$ collectively denotes the matter fields.
There is an extensive literature on theories of gravity of this type; see the review articles
\cite{Hehl:1976kj,Gronwald:1997bx,Shapiro:2001rz,Hammond:2002rm}.

It is often useful to re-express these theories using the tetrad formalism \cite{weinberg}.
In this formalism the independent variables are taken to be a tetrad
of four linearly independent vector fields $e^{\ \mu}_{a}(x)$,
and a tetrad connection $\omega_\mu^{\ ab} = - \omega_\mu^{\ ba}$ defined by
\be\label{tetconn}
{\vec \nabla}_{{\vec e}_{a}} {\vec e}_{b} = e^{\ \mu}_{a} \, \omega_{\mu\ b}^{\ c} \, {\vec e}_{c}.
\ee
Here tetrad indices $a$ run from 0 to 3 and are raised and lowered
using the Minkowski metric $\eta_{ab} \equiv {\rm diag}(-1,1,1,1)$.
The two sets of variables, $g_{\mu\nu}, \Gamma^\lambda_{\ \mu\nu}$ and
$e^{\ \mu}_{a}, \omega_\mu^{\ ab}$ are related by\footnote{The
  connection (\ref{Gammaformula}) is automatically metric compatible
  by virtue of the antisymmetry of $\omega_\mu^{\ ab}$ on $a$ and $b$.}
\bes
\bea
\label{metric}
g_{\mu\nu} &=& \eta_{ab} e_{\ \mu}^{a} e_{\ \nu}^{b}\,, \\
\Gamma^\lambda_{\ \mu\nu} &=& e^{\ \lambda}_{a} \left[ e^{a}_{\
    \nu,\mu} + \omega_{\mu\ b}^{\ a} e^{b}_{\ \nu} \right],
\label{Gammaformula}
\eea
\ees
where the dual basis of one-forms $e^{a}_{\ \mu}$ is defined by
\be
e_{a}^{\ \mu} \, e^{b}_{\ \mu} = \delta^{b}_{a}.
\label{dualbasis}
\ee
The action of the theory in terms of the tetrad variables is
\bea
\label{fullaction0a}
S[e^{\ \mu}_{a},\omega_\mu^{\ ab},\Psi]&=&S_G[e^{\ \mu}_{a},\omega_\mu^{\ ab}]
+S_{\rm matter}[e^{\ \mu}_{a},\omega_\mu^{\ ab},\Psi]. \nonumber \\
\eea
Normally the theory is invariant under local Lorentz transformations
$\Lambda_{a}^{\ b} = \Lambda_{a}^{\ b}(x)$ of the tetrad
\bes
\label{eq:symm}
\bea
\label{eq:symm0}
e^{\ \mu}_{a} &\to& \Lambda_{a}^{\ b} e^{\ \mu}_{b}, \\
\omega_{\mu\ b}^{\ a} &\to& \Lambda^{a}_{\ c} \, \Lambda_b^{\ d} \,
\omega_{\mu\ d}^{\ c} - \Lambda^a_{\ c,\mu} \Lambda_b^{\ c},
\label{eq:symm1}
\eea
\ees
together with the corresponding transformations of any
fermionic matter fields.

There are three different categories of theories involving torsion:
\begin{itemize}

\item Theories in which torsion is an independent dynamical variable
  and the field equations for torsion are
  algebraic, for example the Einstein-Cartan theory \cite{Hehl:1976kj}
  in which the gravitational action is proportional to the Ricci scalar.
  In these theories the torsion vanishes in vacuum.

\item Theories in which the torsion tensor is an independent dynamical
  variable and a propagating degree of freedom, for example Refs.\
  \cite{Hehl:1976kj,Neville:1979rb,Sezgin:1979zf,Sezgin:1981xs,Carroll:1994dq,Shapiro:2001rz,Hammond:2002rm}.
  The source for torsion is the tensor
$$
\sigma_{ab}^{\ \ \mu} \equiv \frac{1}{\sqrt{g}} \frac{ \delta S_{\rm
    matter}}{\delta \omega_\mu^{\ ab}}.
$$
For the standard, minimal coupling of torsion to matter, this tensor
is a measure of density
of fundamental or intrinsic spin, and thus is very small when averaged
over macroscopic distances in unpolarized matter
\cite{Hehl:1976kj,Shapiro:2001rz}.  (For non-standard couplings it is
conceivable that this tensor could be non-negligible.)

\item Theories in which the torsion is not an independent dynamical
  variable, but is specified in terms of some other degrees of freedom
  in the theory, for example a scalar potential
  \cite{Hojman:1978yz,Hojman:1979mg} or a rank 2
  tensor potential \cite{Gruver:2001tt,Hammond:2002rm,Kleinert:1998cz}.

\end{itemize}

A particular special case of the third category are the so-called {\it
  teleparallel theories} \cite{Hehl:1978yt,Hayashi:1979qx,Kopczynski:1982,Hayashi:1981,
Mueller-Hoissen:1983vc,Mueller-Hoissen:1984wq,Nester:1988,Cheng:1988zg,
Blagojevic:2000qs,Blagojevic:2000pi,Obukhov:2002tm,Maluf:2003fs,Mielke:2004gg,
Obukhov:2004hv,Leclerc:2004uu,Leclerc:2005jr}.  In these theories the only dynamical variable
is the tetrad $e_a^{\ \mu}$, the tetrad connection $\omega_\mu^{\ ab}$ is not an independent variable.
In addition the local Lorentz transformations (\ref{eq:symm0}) are {\it not} a symmetry of the theory.
The tetrad therefore contains 6 extra physical degrees of freedom
which are normally gauged away by the local Lorentz symmetry.  In
linear perturbation theory about flat spacetime these extra
degrees of freedom act like a antisymmetric, rank 2 tensor potential
for the torsion; see Sec.\ \ref{linear} below for more details.
In teleparallel theories the torsion is defined to be
\begin{equation}
\label{torsion00}
S_{\mu\nu}^{\ \ \lambda} = \frac{1}{2}e_{a}^{\ \lambda} (e^{a}_{\ \nu,\mu}- e^{a}_{\ \mu,\nu} )\,.
\end{equation}
The form of this equation is invariant under coordinate transformations but not
 under the local Lorentz transformations (\ref{eq:symm0}).
It follows from this definition and from
Eq.\ (\ref{Gammaformula}) that the tetrad connection
$\omega_\mu^{\ ab}$ vanishes, and it follows that the Riemann tensor
\bea
R^{ab}_{\ \ \mu\nu} &\equiv& e^a_{\ \lambda} e^b_{\ \sigma}
R^{\lambda\sigma}_{\ \ \,\mu\nu} \nonumber \\
&=&
 \omega_{\nu\ \ ,\mu}^{\ ab} -  \omega_{\mu\ \ ,\nu}^{\ ab}
+\omega_{\mu\ c}^{\ a} \, \omega_{\nu}^{\ cb}
-\omega_{\nu\ c}^{\ a} \, \omega_{\mu}^{\ cb}\ \ \ \ \
\eea
also vanishes.
Thus in teleparallel theories the curvature
(\ref{eq:Riemann1}) of the full connection
vanishes, and so on the right hand side of Eq.\ (\ref{eq:Riemann1})
the contorsion terms must cancel the curvature term.
Hence, if the spacetime metric is close to that predicted by
general relativity, so that ${\tilde R} \sim M/r^3$ at a distance $r$ from a
mass $M$, then the torsion must be of order $S \sim K \sim M/r^2$.
Therefore, teleparallel theories generically predict a non-negligible
torsion for macroscopic, unpolarized bodies, unlike conventional
torsion theories.

\subsection{Constraining torsion with Gravity Probe B}

The prevailing lore about torsion theories has been that they are very difficult
to distinguish from general relativity, since the torsion generated by
macroscopic bodies is normally negligibly small for the reasons
discussed above \cite{Stoeger:1985,Hammond:2002rm}.
However, a recent paper by Mao, Tegmark, Guth and Cabi (MTGC) \cite{mtgc}
points out that teleparallel theories are an exception in this
regard.  They suggest that Gravity Probe B (GPB) might be an ideal tool to
probe such torsion theories.  In particular they argue that since
the angular momentum of macroscopic bodies generates torsion, one
would expect that the angular momentum of test bodies such as the GPB
gyroscopes would couple to the Earth's torsion field, by the
principle of ``action equals reaction''.

MTGC also review the literature on the equations of motion and spin precession of test bodies
in torsion theories \cite{Hojman:1978wk,Stoeger:1979,Hojman:1979mg,Yasskin:1980bu,
Cognola:1981cm,Kopczynski:1986ep,Hayashi:1990nw,Nomura:1991,Kleinert:1996yi,Kleinert:1998as,
Kleinert:1998cz,Arcos:2004ig}.  They argue that because there is some
disagreement in this literature, and because the precise form of the coupling of torsion
to matter is not known, it is
reasonable to assume that test bodies fall along geodesics of the full
connection (called autoparallels), and to assume that the spin of a gyroscope is
parallel transported with respect to the full connection.
They introduce a theory of gravity called the
Einstein-Hayashi-Shirafuji (EHS) theory, a generalization of an
earlier teleparallel theory of Hayashi and Shirafuji
\cite{Hayashi:1979qx}, and compute the constraints that GPB will be
able to place on this theory for their assumed equations of motion and
spin transport.

In this paper we re-examine the utility of GPB as a probe of torsion
gravity theories.  We agree with the general philosophy expressed by MTGC that
the precise form of the coupling of torsion to matter is something that should
be tested experimentally rather than assumed a priori.
However, while it is conceivable that there
could exist couplings which would predict a detectable torsion signal
for GPB, we show that teleparallel theories and the EHS theory in particular do
not.

We start by discussing the action equals reaction principle.
This appears to be a robust and very generic argument, indicating that the angular momentum of a
test body should couple to torsion in theories where spinning bodies
generate torsion.  However, in fact there is a loophole in the
argument, and in particular it does
not apply to the EHS theory, as we
show in detail in Sec.\ \ref{revised} below.  The
nature of the loophole
can be understood using a simple model.
Consider in Minkowski spacetime the following theory of two scalar fields $\Phi_1$
and $\Phi_2$ and a particle of mass $m$
\bea
S &=&- \frac{1}{2} \int d^4 x \left[  (\nabla \Phi_1)^2 +(\nabla \Phi_2)^2
\right] \nonumber \\
&& - \int d\lambda (m + q \Phi_1) \sqrt{ - \eta_{\mu
\nu}
\frac{d x^\mu}{d\lambda} \frac{d x^\nu}{d\lambda} }.
\eea
Here $\lambda$ is a parameter along the worldline and $q$ is a scalar charge.
In this theory the particle generates a $\Phi_1$ field but not a $\Phi_2$
field, and correspondingly it feels a force from the $\Phi_1$ field but
not the $\Phi_2$ field, in accordance with the action equals reaction
idea.  However, consider now the theory in non-canonical variables
${\tilde \Phi_1} = \Phi_1$, ${\tilde \Phi_2} = \Phi_1 + \Phi_2$.
In terms of these variables
the particle generates both a ${\tilde \Phi_1}$ field and a ${\tilde
  \Phi_2}$ field, but feels a force only from the ${\tilde \Phi_1}$
field, in violation of action equals reaction.
Thus, we see that the action equals reaction principle can only be
applied to the independent dynamical variables in the theory, which
diagonalize the kinetic energy term in the action.  In the EHS theory,
the torsion and metric are not independent dynamical variables; see
Sec.\ \ref{revised} below.


\subsection{The Einstein-Hayashi-Shirafuji theory}

In the remainder of this paper we examine in detail the EHS theory suggested by MTGC.
In this theory the defining relation (\ref{torsion00}) between torsion and
tetrad for teleparallel theories is replaced by
\begin{equation}
\label{torsion}
S_{\mu\nu}^{\ \ \lambda} = \frac{\sigma}{2}e_{a}^{\ \lambda} (e^{a}_{\ \nu,\mu}- e^{a}_{\ \mu,\nu} )\,,
\end{equation}
where parameter $\sigma$ lies in the range $0 \leq \sigma \leq 1$.
For $\sigma=1$ this reduces to the teleparallel case, while for
$\sigma=0$ the torsion tensor vanishes.  Thus, the EHS theory interpolates
between GR at $\sigma=0$, in which the torsion vanishes but the Riemann tensor is
in general different from zero, and the Hayashi-Shirafuji teleparallel
theory\cite{Hayashi:1979qx} at $\sigma=1$, in which the Riemann tensor
vanishes but the torsion tensor is in general different from zero.

The only dynamical variable in the gravitational sector of this theory
is the tetrad $e_a^{\ \mu}$, and, as for the teleparallel theories, the
theory is generally covariant but not invariant under the local Lorentz
transformations (\ref{eq:symm0}) of the tetrad.
The action for the theory can be written as
\bea
\label{fullaction}
S[e^{\ \mu}_{a},\Psi]&=&S_G[e^{\ \mu}_{a}]
+S_{\rm matter}[e^{\ \mu}_{a},\omega_\mu^{\ ab},\Psi],
\eea
where $\Psi$ denotes the matter fields and, in the second term,
$\omega_\mu^{\ ab}$ denotes the tetrad connection obtained from the
torsion tensor (\ref{torsion})
via Eqs.\ (\ref{kdef}), (\ref{kands}) and (\ref{Gammaformula}).  The
matter action $S_{\rm matter}$ is
not specified by MTGC, so there are different versions of the EHS
theory depending on the form of the coupling to torsion chosen in this
matter action.
The gravitational action is given by
\be
S_G[e^{\ \mu}_{a}] = \int d^4 x \sqrt{-g} \left[ a_1 t_{\mu\nu\lambda}
  t^{\mu\nu\lambda} + a_2 v_\nu v^\nu + a_3 a_\nu a^\nu \right].
\label{gaction}
\ee
Here $a_1$, $a_2$ and $a_3$ are free parameters with dimensions of
mass squared (we use units with $\hbar = c = 1$), and the tensors
$t_{\mu\nu\lambda}$, $v^\nu$ and $a^\nu$ are defined to be the
irreducible pieces of the torsion tensor (\ref{torsion}), but with the factor
of $\sigma$ removed:
\bes
\label{tva}
\bea
v_{\mu} &=& \sigma^{-1} S_{\mu\lambda}^{\ \ \,\lambda}, \\
\label{eq:axialtorsion}
a_{\mu} &=& \frac{1}{6}\sigma^{-1}\epsilon_{\mu\nu\rho\sigma} S^{\sigma\rho\nu}\,,\\
t_{\lambda\mu\nu}&=&\sigma^{-1} S_{\nu(\mu\lambda)}+\frac{1}{6}(g_{\nu\lambda}
v_\mu +g_{\nu\mu} v_{\lambda})-\frac{1}{3}g_{\lambda\mu}v_{\nu}.\ \
\ \ \ \ \ \ \
\eea
\ees
Also in Eq.\ (\ref{gaction}) $g$  denotes the metric determinant,
where $g_{\mu\nu}$  is given in terms of the tetrad by Eq.\ (\ref{metric}).
For fixed $a_1$, $a_2$ and $a_3$, this gravitational action is
independent of the parameter $\sigma$; this parameter enters the
theory only through the dependence of the matter action on the torsion
tensor (\ref{torsion}).\footnote{The parameters defining the theory
  are therefore $(a_1,a_2,a_3,\sigma)$.  MTGC use a different set of
  parameters $(c_1,c_2,c_3,\kappa,\sigma)$ which are not all
  independent.  The relation between the two sets of parameters can be
  derived using the identity of Appendix \ref{sec:identity} and is
$a_1 = \sigma^2 c_1 - 4/(3 \kappa)$,
$a_2 = \sigma^2 c_2 + 4/(3 \kappa)$, $a_3 = \sigma^2 c_3 - 3/ \kappa$.}

In Appendix \ref{sec:identity} we show that the gravitational action
can be rewritten in terms of the Ricci scalar $R(\{\})$ of the
Levi-Civita connection as
\be
S_G[e^{\ \mu}_{a}] = \int d^4 x \sqrt{-g} \left[ d_1 R(\{\})  + 4 d_2
  v_\nu v^\nu + 9 d_3 a_\nu a^\nu \right],
\label{gaction1}
\ee
where $d_1 = - 3 a_1/8$, $d_2 = (a_1 + a_2)/4$, $d_3 = a_3/9 - a_1/4$.
We shall refer to the three-dimensional space parameterized by
$(d_1,d_2,d_3)$ as the gravitational-action parameter space.

As mentioned above, there are different versions of the EHS theory,
depending on the form of the matter action $S_{\rm matter}$ chosen;
MTGC do not specify a matter action.
Consider now what is required in order to predict the
signal seen by GPB.  The experiment consists of an Earth-orbiting satellite carrying four
very stable gyroscopes, and the measured quantity is the time
dependence of the angles between the spins of the gyroscopes and the
direction to a fixed guide star.  To compute this quantity in an
arbitrary Riemann-Cartan spacetime, it is sufficient to know the
equations of motion and of spin transport for a spinning point
particle\footnote{One also needs to know the trajectories of
photons, but these are determined by gauge invariance to be just the null
geodesics of the metric, as in GR \cite{Hehl:1976kj}.}.  These
equations can be computed in principle for any matter action.  MTGC
assume that the matter action is such that the trajectory of the
spinning point particle is either an {\it autoparallel} (a geodesic of the
full connection) or an {\it extremal} (a geodesic of the metric), and
that its spin is parallel transported with respect to the full
connection.

\subsection{Requirements necessary to ensure physical viability of the theory}

\begin{table*}
\noindent
\begin{center}
\begin{tabular}{|l||l|l|l|}
\hline
  & Autoparallel  &  Extremal & Standard
Matter Coupling \\ \hline
Sector of parameter space  &   &  &  \\ \hline
${\cal D}_0 = \left\{ d_2 \ne 0, d_3 \ne 0 \right\}$
& Ghosts \ \ \ \ & Ghosts \ \ \ \ & Ghosts  \\\hline
${\cal D}_1 =\left\{ d_2 \ne 0, d_3 = 0 \right\}$
 & I.V.F.   & I.V.F. &  I.V.F. \\\hline
${\cal D}_2 =\left\{ d_2 = 0, d_3 \ne 0 \right\}$
 & I.V.F.   & I.V.F. &  Consistent
but no GPB torsion signal\ \ \  \\
 &   &  &  (action=reaction violated) \\\hline
${\cal D}_3=\left\{ d_2 = 0, d_3 = 0 \right\}$
 & Inconsistent   & Inconsistent/GR &  Inconsistent \\ \hline
\end{tabular}
\caption{A summary of the status of the Einstein-Hayashi-Shirafuji
  theory \cite{mtgc} in different sectors of its parameter space.  The
  rows of the table are these different sectors; the
  parameters $d_2$ and $d_3$, which appear in the gravitational part of
  the action,
  are defined in Eq.\ (\ref{gaction1}) in the text.
  There are different versions of the Einstein-Hayashi-Shirafuji
  theory depending on the precise form chosen of the coupling of the torsion
  tensor to matter fields.  These different versions are the columns
  of the table.  ``Autoparallel'' means that it is assumed that the
  matter coupling is such that freely falling bodies move on geodesics
  of the full connection, while ``Extremal'' means they move on
  geodesics  of the Levi-Civita connection determined by the metric.
  These were the two cases considered by Mao et al.\ \cite{mtgc}.
  ``Standard Matter Coupling'' means the standard, minimal coupling of
  torsion to matter fields \cite{Hehl:1976kj,Shapiro:2001rz}, which in
  general gives rise to motions of test bodies that is neither
  autoparallel nor extremal.  The meanings of the various entries in
  the table are as follows.  ``Ghosts'' means that some of
  the degrees of freedom in the theory are ghostlike at short distances, 
 signaling an instability that rules
  out the theory.  ``I.V.F.'' means that the theory does
  not have a well posed initial value formulation, and so is ruled
  out.  ``Inconsistent'' means that the theory does not predict the
  value of the torsion tensor, so the motion of test bodies cannot be
  predicted, while ``GR'' means that the theory reduces to general
  relativity.  Finally ``no GPB torsion signal'' means that there is no
  torsion-induced coupling
  between the Earth's angular momentum and that of the Gravity Probe B gyroscopes;
  there is only a coupling between the fundamental spins of the Earth's
  fermions and of those in the gyroscopes, which gives a negligible
   signal as those spins are randomly oriented. }
\label{table:ehs}
\end{center}
\end{table*}

In this paper we constrain the parameter space of the EHS theory by imposing a set of physical requirements.
To simplify the analysis we first linearize the EHS theory with respect to a flat torsion-free spacetime, and then
impose physical requirements on the linearized theory. The linearized theory is completely characterized
by two tensor fields: a symmetric field  $h_{\mu\nu}$, and  an antisymmetric field $a_{\mu\nu}$. In terms of these fields the
 torsion tensor is given by $S_{\mu\nu}^{\ \ \lambda}=\sigma(h^{\lambda}_{\ [\nu,\mu]}-2a^{\lambda}_{\ [\nu,\mu]})/2$, and
the metric is given by   $g_{\mu\nu}=\eta_{\mu\nu}+h_{\mu\nu}$.

We impose three types of requirements.
First, we require that the theory have no ghosts, i.e., that the
Hamiltonian of the theory be bounded from below. 
This requirement rules out most of the three-dimensional parameter
space of the gravitational action $S_G$.
The remaining viable subdomain of the parameter space
consists of two intersecting two-dimensional planes.

Second, we further require that the theory have a well posed initial value formulation.
This means that if the physical degrees of freedom are specified on an
initial spacelike hypersurface,
the future evolution of these degrees of freedom is uniquely determined.
Now, for many theories some of the degrees of freedom are nonphysical
and are associated with a gauge symmetry.
For example, in classical electrodynamics the field equations for the vector potential
$A_\mu$ are invariant under the gauge transformation $A_\mu\rightarrow
A_\mu + \varphi_{,\mu}$.  These field equations therefore do not
predict a unique evolution for the vector potential, and
correspondingly consist of a set of underdetermined partial
differential equations.  Nevertheless, classical electrodynamics has a
well posed initial value formulation, because the degrees of freedom
whose evolution cannot be predicted are pure gauge.

A similar situation arises in linearized EHS theory.
There, the gravitational action is invariant under certain symmetries
of the dynamical variables (not diffeomorphisms), and correspondingly the
field equations form a set of an underdetermined partial differential
equations.
Therefore the theory can have a well posed initial value
formulation only if the undetermined degrees of freedom are pure gauge.
This can be the case only if the matter action is also invariant under
the symmetries.  However, the equations of motion and spin precession
postulated by MTGC do not respect these symmetries, and they
should inherit such a property  from the matter action.  We conclude
therefore
that the initial value formulation is ill-posed.
This argument applies in most of the remaining portion of the parameter
space of the theory.  The subdomain that is not excluded by this
argument and by the requirement of no ghosts consists of
a single line in the three dimensional space.

Third, in this remaining subdomain, the linearized gravitational action reduces to
that of GR; it depends on the tetrad only through the metric
$g_{\mu\nu}$. In particular this means that the torsion
tensor is completely undetermined in vacuum, and so
for a generic matter action, the motion of test bodies
cannot be predicted.  The corresponding inconsistency of the
Hayashi-Shirafuji theory in this limit has been previously discussed
in Refs.\
\cite{Hayashi:1979qx,Nester:1988,Blagojevic:2000qs,Blagojevic:2000pi,Obukhov:2002tm,Mielke:2004gg,Obukhov:2004hv,Leclerc:2004uu}.
This inconsistency is avoided  if one constructs a special matter action
in which  the unpredictability of the torsion tensor is associated  with
a  gauge symmetry of the theory, in this case the  predications of the EHS theory coincide with the predictions of GR.

Finally, our argument that the initial value formulation is ill posed can be
evaded by modifying the coupling of the torsion tensor to matter in
the theory.  Rather than postulating the equations of motion and spin
precession used by MTGC, we instead assume that the coupling of
torsion to matter is the standard, minimal coupling described in
Refs.\ \cite{Hehl:1976kj,Shapiro:2001rz}.  For this coupling, the
matter action {\it is} invariant under the symmetries of the
gravitational action discussed above, in a portion of the parameter
space, and so the theory has a well
posed initial value formulation in which the undetermined degrees of
freedom are interpreted as gauge degrees of freedom.  This
is the interpretation suggested in the original paper by Hayashi and
Shirafuji \cite{Hayashi:1979qx}.
For this case, we again examine the linearized theory for the fields $h_{\mu\nu}$ and $a_{\mu\nu}$.
We find that $h_{\mu\nu}$ satisfies the same equation as the metric perturbation in GR,  while
$a_{\mu\nu}$ satisfies a wave equation (with a suitable choice of
 gauge) whose source is obtained from the intrinsic spin density of matter.  As
  discussed earlier, this implies that for a macroscopic object for which the spins of the
  elementary particles are not correlated over macroscopic scales,
  $a_{\mu\nu}$ will be negligible. 
Hence the spacetime of the linearized theory is
completely characterized by the metric alone, and so its predictions
coincide with those of GR\footnote{This is despite the fact that the
torsion tensor is generically nonzero.  The torsion tensor
is not an independent degree of freedom in this limit, it is given
in terms of the metric by the first term in Eq.\ (\ref{torsionf}).}
and there will be no extra signal in GPB.


A summary of the status of the EHS theory in various different cases discussed
above is given in Table \ref{table:ehs}.

To summarize, there are no cases in which the EHS theory gives a
detectable torsion signal in GPB.  However, it is nevertheless possible that
other torsion theories in the other categories discussed in Sec.\
\ref{sec:torsiontheories}, with suitable choice of matter coupling,
could predict a detectable signal.  Various possibilities for
non-minimal couplings are discussed by Shapiro
\cite{Shapiro:2001rz}.  It would be interesting to find
a torsion theory that predicts a detectable torsion signal for GPB;
such a theory would be an example to which the theory-independent
framework developed by MTGC (a generalization of the parameterized post-Newtonian
framework to include torsion) could be applied.

\subsection{Organization of this paper}

This paper is organized as follows.
In Sec.\ \ref{linear} we derive the dynamical variables and the action
of the linearized EHS theory.
In Sec.\ \ref{antis} we study the action for the antisymmetric field
$a_{\mu\nu}$, temporarily setting the symmetric field $h_{\mu\nu}$ to zero.
We show that this theory has  ghosts on a subdomain
of the parameter space. 
Section \ref{completeact} and Appendix \ref{sec:ghost1} extend this result to
the complete linearized theory, including the symmetric field $h_{\mu\nu}$,
thereby ruling out a subdomain of the parameters space.
We then focus on the complementary subdomain and show that it is invariant under certain
symmetries.  Section \ref{init} reviews the necessary requirements
for a well posed  initial-value formulation.  In Sec.\ \ref{obs} we use
these requirements to rule out a subdomain of the parameter space that has an ill-posed
initial value formulation.  The remaining portion of the parameter space is discussed in Sec. \ \ref {consehs}

Finally, section \ref{revised} considers the EHS theory with the standard matter-torsion coupling.
In a certain portion of parameter space this theory has a
well-posed initial value formulation and no 
ghosts, but we show that the deviations of its predictions
from those of GR are negligible for unpolarized macroscopic bodies. 
Final conclusions are given in Sec.\ \ref{conc}.

\section{Linearization about flat, torsion-free spacetime of the
  Einstein-Hayashi-Shirafuji theory}\label{linear}

\subsection{Action and variables of linearized theory }

To linearize the EHS theory we first decompose the tetrads
$e_a^{\ \mu}$ and dual one-forms $e^a_{\ \mu}$
into background tetrads and one-forms and perturbations:
\bes
\label{dectetrad}
\bea
e_{a}^{\ \mu} &=& b_{a}^{\ \mu}+\delta c_{a}^{\ \mu}, \\
e^{a}_{\ \mu} &=& b^{a}_{\ \mu}+\delta e^{a}_{\ \mu}.
\eea
\ees
We assume that the background tetrads $b_{a}^{\ \mu}$ are constants
for which the metric (\ref{metric}) is the Minkowski metric,
$g_{\alpha\beta}=\eta_{\alpha\beta}$, and for which the torsion
(\ref{torsion}) is vanishing, $S_{\mu\nu}^{\ \  \lambda}=0$.
Thus, to zeroth order, the spacetime is flat and torsion-free.
Throughout this paper we will work to leading order in the tetrad perturbations.
Hereafter, unless we explicitly state otherwise, Greek indices are
raised and lowered with $\eta_{\mu\nu}$ and Latin indices with $\eta_{ab}$.

From the definition (\ref{dualbasis}) of the dual basis applied to both the full tetrads and the background
tetrads, we find that $\delta e^{a}_{\ \mu}b_{a}^{\ \nu}=-\delta
c_{a}^{\ \nu}b^{a}_{\ \mu}$.  Thus we can take $\delta e^a_{\ \mu}$ to be the
fundamental variable of the theory.
We next convert this quantity into a spacetime rank 2 tensor using the
background tetrad, and take the independent symmetric and
antisymmetric pieces.  This yields the definitions
\bes
\label{decomp}
\bea
h_{\mu\nu} &\equiv& 2 \delta e^{b}_{\ (\mu} b^{a}_{\ \,\nu)} \eta_{ab}, \\
a_{\mu\nu} &\equiv& \delta e^{b}_{\ [\mu} b^{a}_{\ \,\nu]} \eta_{ab}.
\eea
\ees
The formulae (\ref{metric}) and (\ref{torsion}) for the metric and
torsion now yield
\bes
\label{gs}
\bea
g_{\mu\nu} &=& \eta_{\mu\nu}+  h_{\mu\nu}, \\
S_{\mu\nu}^{\ \ \lambda} &=& \frac{\sigma}{2} \left( h^{\lambda}_{\
    [\nu,\mu]}-2a^{\lambda}_{\ [\nu,\mu]}\right).
\label{torsionf}
\eea
\ees
The linearized gravitational action $S^{\rm linear}_G $ can now be
obtained by substituting the expressions (\ref{gs}) for the metric and
torsion into the action (\ref{gaction1}) and expanding to quadratic order in
$h_{\mu\nu}$ and $a_{\mu\nu}$.
The resulting action can be written schematically in the form
\begin{equation}\label{linaction}
S^{\rm linear}_G =    S_S[h_{\mu\nu}]+S_C[h_{\mu\nu},a_{\alpha\beta}]+S_A [a_{\mu\nu}] \,,
\end{equation}
where $S_S$ is a quadratic in the symmetric tensor $h_{\mu\nu}$, $S_A$
is quadratic in the antisymmetric tensor $a_{\mu\nu}$, and $S_C$
contains the cross terms.

\subsection{The antisymmetric term in the action}\label{antis}

We now focus on the antisymmetric term $S_A$ in the action, ignoring
for the moment the other two terms.  For this theory we derive
two kinds of results. First, we constrain the parameter space $(d_1,d_2,d_3)$
by imposing  the requirement of no ghosts, and second 
we derive symmetry transformations under which
the action $S_A$ (with specific parameters) is invariant.
Later in Sec.\ \ref{completeact} we will extend some of these results
to the complete linearized action $S^{\rm linear}_G$.

The antisymmetric action $S_A[a_{\mu\nu}]$ is constructed from
the antisymmetric field $a_{\mu\nu}$ in Minkowski spacetime, and consists of terms that are product of derivatives of
$a_{\mu\nu}$, of the form $(\partial_{\lambda} a_{\mu\nu})^2$.
Actions of this type also arise in
gravitational theories with a non-symmetric metric and have been
extensively studied.  See Refs.\
\cite{VanNieuwenhuizen:1973fi,1993PhRvD..47.1541D} for a discussion of
the existence of ghosts in theories of this type.

There are only three linearly independent terms of the form  $(\partial_{\lambda} a_{\mu\nu})^2$, namely
\[
a_{\mu\lambda}^{\ \ \,,\lambda}a^{\mu\ ,\sigma}_{\ \sigma}\ , \
a_{\mu\nu,\lambda} a^{\mu\nu,\lambda}\ , \ a_{\mu\nu,\lambda} a^{\mu\lambda,\nu}\, .
\]
From these terms one can construct only two functionally independent actions,  since
the identity
\be
a^{\mu\lambda}_{\ \ \,,\lambda} a_{\mu\sigma}^{\ \ \,,\sigma} -
a^{\mu\lambda,\sigma} a_{\mu\sigma,\lambda} = \partial_\lambda (
a^{\mu\lambda} a_{\mu\sigma}^{\ \ \,,\sigma} - a^{\mu\lambda,\sigma} a_{\mu\sigma})
\ee
shows that a linear combination of the terms is a divergence which can
be converted to a surface term upon integrating and thereby discarded.
Therefore,  the most general action of this type can be written as
\begin{equation}\label{la}
S_A=\int d^4 x [d_2 a_{\mu\lambda}^{\ \ \,,\lambda}a^{\mu\sigma}_{\ \ ,\sigma}+ d_3   a_{\mu\lambda}^{*\ \,,\lambda} a^{*\mu\sigma}_{\ \ \ \,,\sigma} ]\,,
\end{equation}
where  $a^{*\mu\nu}\equiv -\epsilon^{\mu\nu\rho\sigma} a_{\rho\sigma}/2$ (here $\epsilon^{\mu\nu\rho\sigma}$ is the Levi Civita tensor of a flat spacetime), and $d_{2}$,$d_{3}$ are free parameters.
Indeed an explicit calculation using Eqs.\ (\ref{tva}), (\ref{gaction1}) and
(\ref{gs}) shows that $S_A$ is given by the expression (\ref{la}),
where the parameters $d_2$ and $d_3$ are those defined after Eq.\
(\ref{gaction1}).
Next, we specialize to a particular Lorentz frame, and rewrite the
action in terms of
the vectors ${\bm E}$ and ${\bm B}$ defined by
$E_i = a_{0i}$ and $B_{i} = \frac{1}{2}\epsilon_{ijk} a_{jk}$, where $i,j,k$
run from $1$ to $3$.
This gives
\begin{equation}\label{lagdivdiv}
a_{\mu\lambda}^{\ \ \,,\lambda}a^{\mu\sigma}_{\ \ ,\sigma}=({\bf \dot{E} +{\bm \nabla}\times B})^2-({\bf {\bm \nabla} \cdot E})^2\,,
\end{equation}
where an overdot denotes differentiation with respect to time. The corresponding expression for
 $a_{\mu\lambda}^{*\ \,,\lambda} a^{*\mu\sigma}_{\ \ \ \,,\sigma}$ has the same form, except for
the substitutions ${\bf E \rightarrow B}$ and ${\bf B \rightarrow
  -E}$.

We now consider the case where both $d_{2}$ and $d_{3}$
 are nonzero. In this case Hamiltonian density corresponding to the
 action (\ref{la}) takes the form
\begin{eqnarray}\label{hamiltonian}
&&{H}_A=\frac{1}{4d_3}({\bm \pi}_{\bf B}+2d_3 {\bm \nabla} {\bf
  \times E})^2+d_3( {\bm \nabla}  {\bf \cdot B})^2-d_3({\bm
  \nabla}{\bf \times E})^2 \nonumber \\
&&+\frac{1}{4d_2}({\bm \pi}_{\bf E}-2d_2 {\bm \nabla}{\bf  \times
  B})^2+d_2({\bm \nabla}{\bf \cdot E})^2-d_2({\bm \nabla}{\bf \times
  B})^2\,, \nonumber \\
\end{eqnarray}
where ${\bm \pi}_{\bf E}$ and ${\bm \pi}_{\bf B}$ denote the momenta conjugate
to ${\bf E}$ and ${\bf B}$.

\subsection{Ghosts}\label{ghostinsa}

We now constrain the values of the  coefficients $d_{2}$ and $d_{3}$ by
 demanding that $\int { H}_A d^3x$ be bounded from below.
Let us start by considering the coefficient $d_3$, and suppose first that $d_3<0$.
At a given point in space, keeping the values of ${\bf E}$, ${\bf B}$ and ${\bm \pi}_{\bf E}$
fixed, we can make ${H}_A$ arbitrarily negative by choosing
${\bm \pi}_{\bf B}$ to be arbitrarily large.  The same is clearly true
for the integrated Hamiltonian $\int H_A d^3x$.
Next suppose that $d_3>0$.
By fixing ${\bf B}$, ${\bm \nabla} \cdot  {\bf E}$, ${\bm \pi}_{\bf E}$ and
adjusting the value of ${\bm \pi}_{\bf B}$ to keep the value of  ${\bm \pi}_{\bf B}+2d_3 {\bf {\bm \nabla} \times E}$  fixed,
we can make ${H}_A$ arbitrarily negative, this time by choosing an arbitrarily large value for ${\bf {\bm \nabla}\times E}$. 
By applying analogous considerations to $d_2$ we reach the conclusion that
the theory defined by $H_A$ has ghosts on the domain 
\be
{\cal D}_0 = \left\{ (d_1,d_2,d_3) \right| \left.  d_2 \ne 0, d_3 \ne 0\right\}
\ee
in parameter space.  Similar analyses can be found in Refs.\
\cite{VanNieuwenhuizen:1973fi,1993PhRvD..47.1541D} for non-symmetric
gravity theories, and in Ref. \cite{KuhfussNitsch} for teleparallel gravity theories.

\subsection{Symmetries}

We now consider the domain in parameter space not excluded by the
above analysis, which consists of the two 2D regions
\bes
\label{defdomain}
\be
{\cal D}_1 = \left\{ (d_1,d_2,d_3) \right| \left.  d_2 \ne 0, d_3 =
  0\right\},
\ee
\be
{\cal D}_2 = \left\{ (d_1,d_2,d_3) \right| \left.  d_2 = 0, d_3 \ne
  0\right\},
\label{calD2def}
\ee
together with the line
\be
{\cal D}_3 = \left\{ (d_1,d_2,d_3) \right| \left.  d_2 = 0, d_3 =
  0\right\}.
\label{calD3}
\ee
\ees
The antisymmetric action $S_A$ vanishes identically on ${\cal D}_3$,
so in this subsection we will not consider ${\cal D}_3$ any further.

We are interested in the symmetries of the antisymmetric action $S_A$
on the domains ${\cal D}_1$ and ${\cal D}_2$.  From the formula
(\ref{la}) for the action we see that these two domains are isomorphic
to one another under the duality transformation
\be
a_{\mu\nu} \to a_{\mu\nu}^* \ \ \ {\rm or} \ \ \ {\bm E} \to {\bm B},
\ {\bm B} \to - {\bm E}.
\ee
Therefore it is sufficient to focus on one of the domains, say ${\cal
D}_1$.  From Eqs.\ (\ref{la}) and (\ref{lagdivdiv}), the antisymmetric
action on this domain is
\be
S_{A\,|{\cal D}_1}=\int d_2[ ({\bf \dot{E}+{\bm \nabla} \times B})^2-({\bf {\bm \nabla} \cdot E})^2]\,d^4 x \,.
\label{actionD1}
\ee

Consider now the initial value problem for ${\bf E}$ and ${\bf B}$.  Suppose that we are
given sufficient initial data on some constant time hypersurface, and that we wish to determine the time evolution of
${\bf E}(t)$ and ${\bf B}(t)$ using the action
(\ref{actionD1}).  Note that this action is independent of the
longitudinal part of ${\bf B}$, which means
that the evolution of this longitudinal part  can be prescribed arbitrarily,
independent of the initial data. Therefore, the field equations for ${\bf E}$ and ${\bf B}$ must form a set of an
underdetermined partial differential equations.
As discussed in the introduction, this can only be consistent if the
undetermined degrees of freedom can be interpreted as being gauge
degrees of freedom.

The action (\ref{actionD1}) is invariant under the symmetry
\begin{equation}\label{sym2}
a_{\mu\nu}\rightarrow a_{\mu\nu}+\epsilon_{\mu\nu\alpha\beta}\chi^{\alpha,\beta}\,,
\end{equation}
where $\chi^\alpha(x)$ is an arbitrary vector field.  This can be seen
from the fact that the action (\ref{actionD1}) is given by the first
term in Eq.\ (\ref{la}), which depends on $a_{\mu\nu}$ only through
its divergence $a_{\mu\nu}^{\ \ \,,\nu}$.  Similarly on the domain
${\cal D}_2$ the antisymmetric action (\ref{la}) is invariant under
the symmetry
\begin{equation}\label{sym3}
a_{\mu\nu}\rightarrow a_{\mu\nu}+\chi_{[\mu,\nu]}.
\end{equation}
These are the symmetries that are responsible for the indeterminacy in
the evolution equations.  We will study in later sections the conditions
under which these symmetries can be interpreted as gauge symmetries,
thus allowing the theory to have a well posed initial value
formulation.
As discussed in the introduction, the gauge symmetry
interpretation requires the matter action to be invariant\footnote{A
  general discussion of the
problems that arise when the gravitational action is
invariant under a symmetry not shared by the matter action
can be found in Leclerc \cite{Leclerc:2004uu}.}
under the
symmetries (\ref{sym2}) and (\ref{sym3}).

\subsection{The complete linearized action}\label{completeact}

Up to now we have ignored the pieces $S_S$ and $S_C$ of the complete
linearized action (\ref{linaction}), and have studied
a reduced theory depending only on the antisymmetric field
$a_{\mu\nu}$ described by the action $S_A$ alone. We showed that this
reduced theory has ghosts if both $d_2$ and $d_3$ are
nonzero, i.e., on the domain ${\cal D}_0$.
In Appendix \ref{sec:ghost1}
this result is generalized to the complete linearized theory,
including the symmetric field $h_{\mu\nu}$,
showing that the complete theory also has ghosts in the domain ${\cal
  D}_0$.
Essentially we show that
that whenever the Hamiltonian
$\int H_A d^3x$ is unbounded from below, then the corresponding
Hamiltonian of the full linearized theory is also unbounded from below.

The symmetries (\ref{sym2}) and (\ref{sym3}) of the reduced theory on the
domains ${\cal D}_1$ and ${\cal D}_2$ also generalize to the complete
linearized theory.  This can be seen as follows.  Since the symmetries
only involve the antisymmetric field $a_{\mu\nu}$, the only
additional term in the complete action (\ref{linaction}) whose
invariance needs to be checked is the cross term $S_C[h_{\mu\nu}, a_{\rho\sigma}]$.
This cross term can be written as
\[
S_C=\int  h_{\alpha\beta,\gamma}a_{\mu\nu,\rho}P^{\alpha\beta\gamma\mu\nu\rho}\,d^4x\,,
\]
where $P^{\alpha\beta\gamma\mu\nu\rho}$ is a tensor constructed from the Minkowski metric.
Integrating by parts and discarding a surface term yields
\[
S_C=-\int  h_{\alpha\beta}a_{\mu\nu,\rho\gamma} {P}^{\alpha\beta\gamma\mu\nu\rho}\,d^4x\,.
\]
At least two of the indices on $a_{\mu\nu,\rho\gamma}$ must be
contracted with one another, and since $a_{\mu\nu}$ is antisymmetric
it follows that only divergence terms of the
form $a_{\mu\sigma,\ \gamma}^{\ \ \ \sigma}$ can appear.  These
divergence terms are invariant under the symmetry (\ref{sym2}).

For the symmetry (\ref{sym3}), we compute explicitly the cross term $S_C$
specialized to the domain ${\cal D}_2$.
Since $d_2=0$, the only term in the general action (\ref{gaction1}) that
can contribute to this cross term is the $a_\nu a^\nu$ term involving the square of the
axial piece of the torsion; the Ricci scalar term
depends only on $h_{\mu\nu}$.
Using the definition (\ref{eq:axialtorsion})
of this axial piece together with the formula (\ref{torsionf}) for the
torsion in terms of $h_{\mu\nu}$ and $a_{\mu\nu}$ gives
\be
a_\mu=\frac{1}{6}\eta_{\mu\lambda}
\epsilon^{\lambda\nu\rho\sigma}a_{\nu\sigma,\rho}.
\label{axialpiece}
\ee
Since this depends only on the antisymmetric field $a_{\mu\nu}$ it
does not generate any cross terms, and we conclude that the cross term
$S_C$ vanishes identically on the domain ${\cal D}_2$.
Therefore the complete action is invariant under the
symmetry (\ref{sym3}) on ${\cal D}_2$.

\subsection{The general relativity limit of the gravitational action}
\label{consehs}

So far we have considered the domains ${\cal D}_0$, ${\cal D}_1$ and
${\cal D}_2$ of the gravitational action parameter space.  We now
focus on the remaining domain ${\cal D}_3$.
From the definitions (\ref{calD3}) and (\ref{gaction1}) we find that
in this domain the gravitational action reduces to that of general
relativity, so it is invariant under local Lorentz transformations of the tetrad
$e_a^{\ \mu}\rightarrow\Lambda_{a}^{\ b}(x)e_{b}^{\ \mu}$.
This invariance  guarantees that the left hand side of the  Euler-Lagrange equation
\[
-\frac{\delta S_G}{\delta e^{a}_{\ \rho}}e^{a}_{\ \mu}g_{\rho\nu}=\frac{\delta S_{\rm matter} } {\delta e^{a}_{\ \rho}}e^{a}_{\ \mu}g_{\rho\nu},
\]
is  a symmetric tensor \cite{weinberg}. Consistency now requires  the  right hand side to be symmetric.
However, the torsion tensor is not invariant under local Lorentz transformations. Therefore,
a generic matter action that couples between the torsion tensor and  matter fields would  break the local Lorentz symmetry. This matter action
 produces a nonsymmetric right hand side for the   Euler-Lagrange equation, thereby rendering the theory inconsistent.
The corresponding inconsistency of the
Hayashi-Shirafuji theory in this limit has been previously discussed
in Refs.\
\cite{Hayashi:1979qx,Nester:1988,Blagojevic:2000qs,Blagojevic:2000pi,Obukhov:2002tm,Mielke:2004gg,Obukhov:2004hv,Leclerc:2004uu}.
The inconsistency is avoided  if the  matter action is invariant under local Lorentz transformations. In this case the
torsion tensor is undetermined by the EHS theory, and the theory reduces to GR.

\section{Initial value formulation of the theory}\label{init}

In this section we focus on the domains ${\cal D}_1$ and ${\cal
D}_2$ that are not ruled out by the existence of ghosts, 
and examine in more detail the conditions under which
the theory on these domains has a well posed initial value
formulation.

Suppose that we specify as initial data  $\{a_{\mu\nu},h_{\alpha\beta}\}$ and
$\{\dot{a}_{\mu\nu},\dot{h}_{\alpha\beta}\}$ on some initial constant
time hypersurface, and ask whether the evolution of the fields for all
subsequent times is uniquely determined.
Now the action (\ref{la})
is invariant under certain symmetries which allow us to generate new solutions
that correspond to the same initial data.
These symmetries consist of, first, diffeomorphisms
$x^\mu\rightarrow x^\mu - \xi^\mu(x)$ under which the fields
transform as
\bes
\label{diff}
\begin{eqnarray}
&&h_{\mu\nu}\rightarrow h_{\mu\nu}+2\xi_{(\mu,\nu)}\,,\\
&&a_{\mu\nu}\rightarrow a_{\mu\nu}+\xi_{[\nu,\mu]}\,,
\end{eqnarray}
\ees
and second, the symmetries (\ref{sym2}) on ${\cal D}_1$ and
(\ref{sym3}) on ${\cal D}_2$.  By invoking one of these transformations
in a spacetime region to the future of the initial data hypersurface,
we can generate new solutions for the field-equations that correspond
to the same initial data.  Therefore the evolution of the fields $h_{\mu\nu}$ and
$a_{\mu\nu}$ cannot be uniquely predicted.
For the theory to have a well-posed initial value formulation it is
necessary that {\em all} of these transformations correspond to gauge
symmetries, which
means that all observables should remain invariant under these transformations.
Then the failure of the theory to uniquely predict
$h_{\mu\nu}$ and $a_{\mu\nu}$ is merely associated with the
unpredictability of nonphysical degrees of freedom.

We now focus on the observables that will be measured by the GPB
experiment.  If we use the equations of motion and spin
transport for test bodies postulated by MTGC, then these observables
are not invariant under the symmetries (\ref{sym2}) and (\ref{sym3}),
as we now show.
Thus the initial value formulation is ill posed for these postulated
equations of motion.

\subsection{Observations with Gravity Probe B}
\label{obs}


We focus on one of the four GPB gyroscopes, and represent it as
a particle with trajectory $z^\alpha(\tau)$ where $\tau$ is proper time, and with spin $s^\alpha(\tau)$.
We let the 4-momentum of the photons from the distant fixed guide star be $k^\alpha$.
Let $\theta$ be the angle, as measured in the frame of the gyroscope, between its spin and the direction
to the guide star.  Then we have
\be
\cos \theta = \frac{ {\vec s} \cdot {\vec k} }{ ({\vec u} \cdot {\vec k} )\sqrt{ {\vec s}^2 }},
\label{deftheta}
\ee
where $ {\vec s} \cdot {\vec k}=g_{\mu\nu} s^\mu k^\nu$.
This can be seen from the formulae for these vectors in the rest frame
of the gyroscope: ${\vec u} = (1,\bf{0})$, ${\vec k} = \omega(1,-{\bf n})$
and ${\vec s} = (0,{\bf s})$, where ${\bf n}$ is a unit vector in the
direction of the star.


The equations of motion and spin precession postulated by MTGC are
\begin{equation}\label{motion}
\frac {D u^\mu}{D \tau}=0\ , \  \frac{D s^\mu}{D \tau}=0\,,
\end{equation}
where $u^\mu=d z^\mu / d\tau$ is the 4-velocity and
\be
\frac{D}{D \tau} \equiv u^\mu \nabla_\mu
\ee
is the covariant derivative operator along the worldline with respect to the full connection $\nabla_\mu$.
In other words, the particle travels on a geodesic of the full
connection (an autoparallel curve).
MTGC also consider the possibility that the equation of motion is
\begin{equation}\label{extermal}
\frac {\tilde{D} u^\mu}{{\tilde D} \tau} \equiv u^\nu {\tilde \nabla}_\nu u^\mu =0,
\end{equation}
where ${\tilde \nabla}_\nu$ is the Levi-Civita connection determined
by the metric, so that the
particle travels along a geodesic of the metric (an extremal curve).
As mentioned by MTGC  this possibility is theoretically inconsistent since  by
Eqs. (\ref{motion}) the orthogonality of $u^\mu$ and $ s_\mu $ is not maintained
during the evolution. Nevertheless, we shall also consider this possibility below.
Finally, as mentioned above, photons follow null geodesics of the metric:
\be
k^\mu {\tilde \nabla}_\mu k^\nu =0.
\label{light}
\ee

We now apply the operator $D/D\tau$ to the formula (\ref{deftheta}) for the
angle $\theta$, and use the autoparallel equations of motion (\ref{motion}) together with
$\nabla_\mu g_{\mu\lambda}=0$.  This gives
\begin{equation}\label{rate}
\frac{d [\cos \theta ]}{d\tau}= \frac{ 1 }{ ({\vec u} \cdot {\vec k}
  )\sqrt{ {\vec s}^2 }}
\left[ \left( {\vec s} - \frac{ {\vec s} \cdot {\vec k} }{{\vec u}
      \cdot {\vec k}} {\vec u} \right) \cdot \frac{ D {\vec k} }{D
    \tau} \right].
\end{equation}
The measurable
accumulated change in $\cos(\theta)$ in the interval $\tau_1\rightarrow\tau_2$
 is therefore
\begin{equation}\label{total}
\Delta [\cos(\theta)]=\int_{\tau_1}^{\tau_2} d\tau \frac{ 1 }{ ({\vec u} \cdot {\vec k}
  )\sqrt{ {\vec s}^2 }}
\left[ \left( {\vec s} - \frac{ {\vec s} \cdot {\vec k} }{{\vec u}
      \cdot {\vec k}} {\vec u} \right) \cdot \frac{ D {\vec k} }{D
    \tau} \right].
\end{equation}

Next, we examine how the change in angle (\ref{total}) transforms
under the symmetry transformations (\ref{sym2}) and (\ref{sym3}). For
this purpose it is sufficient to consider the motion of the gyroscope in
a flat, torsion-free spacetime, since we are working to linear order.
We use Lorentzian coordinates where  $\biglb\{
\stackrel{\mu}{\scriptstyle \alpha\beta} \bigrb\}=\Gamma^{\mu}_{\
  \alpha\beta}=0$, which implies that for an initially static gyroscope $u^\mu$,
$s^\mu$ and $k^\mu$ are all constants, so that
$\Delta [\cos(\theta)]=0$.

Consider now the effect of the transformations (\ref{sym2}) or
(\ref{sym3}).
Denoting the transformed quantities with primes
we find that $\biglb\{ \stackrel{\mu}{\scriptstyle \alpha\beta} \bigrb\}'=\biglb\{ \stackrel{\mu}{\scriptstyle \alpha\beta} \bigrb\}=0$,
$\Gamma^{'\alpha}_{\ \ \mu\nu}=0+\delta \Gamma^{\alpha}_{\ \mu\nu}$.
From Eqs.\ (\ref{motion}) we find that  $z'^{\mu}(\tau')= z^\mu(\tau) + \delta z^\mu(\tau)$ and
$s'^{\mu}(\tau') = s^\mu(\tau)+\delta s^\mu(\tau)$, where both $\delta z(\tau)$ and $\delta s^\mu$
are $O(\chi)$. It turns out that the precise expressions for these quantities
are not required for our calculation.  Notice that for
a fixed distant star the field $k^\mu(x)$ (before the transformation)
is approximately constant in a neighborhood of the gyroscope, in the sense that $k^\mu_{\ \, ,\nu}=0$.
Now, Eq.\ (\ref{light}) implies that $k^\mu(x)$ remains invariant
under the transformations.
Therefore the derivative of
$k'^\mu(x)$ along $z'(\tau')$ is given by
\be
\frac{D k'^{\mu}}{D\tau'}=\delta\Gamma^\mu_{\ \alpha\beta}
u'^{\alpha} k'^{\beta}.
\label{dkdt}
\ee
Substituting Eq.\ (\ref{dkdt}) and $s'_\mu(\tau')$
into   Eq.\ (\ref{rate}), and retaining only
  terms which are $O(\chi)$
(so we can drop the distinction between $\tau$ and $\tau'$) we obtain
\bea
\label{transrate}
\frac{d [\cos(\theta')]}{d\tau}&=&
\frac{ 1 }{ ({\vec u} \cdot {\vec k}
  )\sqrt{ {\vec s}^2 }}
\bigg[ s_\mu \delta \Gamma^\mu_{\ \alpha\beta} u^\alpha k^\beta  \nn \\
&&- \frac{ {\vec s} \cdot {\vec k}}{{\vec u} \cdot {\vec k}} \delta
\Gamma^\mu_{\ \alpha\beta} u_\mu u^\alpha k^\beta
\bigg].
\eea
From Eqs.\ (\ref{kdef}), (\ref{kands}) and (\ref{gs})
the change $\delta\Gamma^\mu_{\ \alpha\beta}$ in the
connection coefficients is
\be
 \delta\Gamma^\mu_{\ \alpha\beta} =\sigma \eta_{\beta\lambda}\epsilon^{\mu\lambda\rho\sigma} \chi_{\sigma,\rho\alpha}\,,
\ee
for the symmetry (\ref{sym2}) and
\be
\delta\Gamma^\mu_{\ \alpha\beta} =(\sigma/2) (\chi_{\beta,\alpha}^{\ \
  \ \ \mu}-\chi^\mu_{\ ,\alpha\beta})
\ee
for the symmetry (\ref{sym3}).

We substitute these transformation rules into Eq.\ (\ref{transrate}) and then substitute the result into Eq.\ (\ref{total}).
Recalling that $\chi_\alpha$ are arbitrary functions of the coordinates,
we find  that by invoking the transformations (\ref{sym2}) or (\ref{sym3}) we can
set $\Delta [\cos(\theta)]$ to have an arbitrary value.
Thus, the observable angle is not invariant under the symmetry
transformations, and hence they cannot be interpreted to be gauge
transformations.  [In particular this implies that the matter action
must be non-invariant].  Consequently the initial value formulation of
the theory is ill posed.

We now repeat this analysis for extremal worldlines satisfying Eq.\ (\ref{extermal}).
Equation (\ref{transrate}) acquires an additional term
\be
- \frac{ {\vec s} \cdot {\vec k} }{ ({\vec u} \cdot {\vec k}
  )^2\sqrt{ {\vec s}^2 }} {\vec k} \cdot \frac{D {\vec u}}{D\tau}
 =
- \frac{ {\vec s} \cdot {\vec k} }{ ({\vec u} \cdot {\vec k}
  )^2\sqrt{ {\vec s}^2 }}
\left( \delta \Gamma^\mu_{\ \alpha\beta} k_\mu u^\alpha u^\beta
\right) \nn
\ee
on the right hand side.  In addition the
change to
extremal worldlines alters the quantity $\delta u^\mu$, but this does not appear in the final
formula (\ref{transrate}).  As before we find that the initial value formulation is ill posed.

Finally, MTGC also consider the possibility of an extremal worldline together with the following equation for the spin precession
\be\label{stensor}
\frac{D s_{\alpha\beta}}{D \tau}=0.
\ee
Here the antisymmetric tensor $s_{\alpha\beta}$ is
related to the particle spin through $s^\mu=\epsilon^{\mu\nu\rho\sigma} u_{\nu} s_{\rho\sigma}$.
This relation guarantees that the orthogonality
condition $s^\mu u _{\mu}=0$ is satisfied throughout the motion of the particle.\footnote{
As noted by MTGC, normally one also demands a ``transversality''  condition $s_{\alpha\beta}u^\beta=0$.
However, by Eqs. (\ref{extermal}) and (\ref{stensor}) this condition can not be maintained along an extremal worldline.
This also implies that the norm of $s^\mu$ is not constant along the evolution.}
By examining the transformation of $\Delta[\cos(\theta)]$ under the symmetries  (\ref{sym2})
and (\ref{sym3}) for  an extremal worldline for which
 the law for the spin precession  is given by (\ref{stensor}),
 we find as before that the initial value formulation is ill posed.

\section{Einstein-Hayashi-Shirafuji theory with standard matter coupling}
\label{revised}

The analysis so far suggests that in order to obtain a consistent theory,
one needs to choose a matter action which is invariant under the
symmetries (\ref{sym2}) or (\ref{sym3}) of the gravitational action.
This would allow those symmetries to be interpreted as gauge
symmetries and allow the theory to have a well posed initial value
formulation.  In this section we show that the standard, minimal
coupling of matter to torsion \cite{Hehl:1976kj,Shapiro:2001rz} does
respect the symmetry (\ref{sym3}), and so one does obtain a consistent
linearized theory on the domain ${\cal D}_2$
by choosing this coupling.\footnote{Note that this implies in
particular that the standard matter coupling does not predict either
extremal curves or autoparallel curves for the motions of test bodies,
since those cases are not invariant under the symmetry
(\ref{sym3}).}
For the special case of $\sigma=1$, this matter action
is the one suggested in the original Hayashi-Shirafuji paper \cite{Hayashi:1979qx}.
Following the analysis of Hayashi and Shirafuji, we show that
the predictions of this linearized theory coincide with those of
linearized GR for macroscopic sources with negligible net intrinsic spin.

The standard Dirac action in an Einstein-Cartan spacetime
is
$S_D=\int  d^4x \sqrt{-g} { L}_D$, where
\bes
\bea
\label{ld1}
{L}_D &=&  \frac{i}{2}  e_{a}^{\ \mu} ( \bar{\psi} \gamma^{a} D_{\mu}
\psi-\overline{D_{\mu}\psi}\gamma^{a}{\psi})-m\bar{\psi}\psi \,,\
\ \ \  \\
\label{Dmudef}
D_{\mu} &=& \partial_{\mu}
-\frac{i}{4}\sigma^{bc}g_{\rho\nu}e_{b}^{\ \nu}
\nabla_{\mu}e_{c}^{\ {\rho}}, \\
\nabla_{\mu}e_{c}^{\ \rho} &=& \partial_{\mu} e_{c}^{\
  \rho}+\Gamma^{\rho}_{\ \mu\nu}  e_{c}^{\ \nu}, \\
\sigma^{bc} &= &\frac{i}{2} [\gamma^{b},\gamma^{c}],
\eea
\ees
and $\gamma^{b}$ are Dirac matrices with the representation given in
Ref. \cite{bd} that satisfy $\gamma^{a}\gamma^{b}+\gamma^{b}\gamma^{a}=-2\eta^{ab}$.
Also $\bar{\psi}$ denotes the adjoint spinor defined by
$\bar{\psi}=\psi^\dagger \gamma^{0}$, where $\dagger$ denotes
Hermitian conjugation.
The torsion tensor enters this action only through the covariant
derivative in Eq.\ (\ref{Dmudef}).
As is well known, this action can be written as
the usual torsion-free
action together with a coupling of the fermion to the axial piece
(\ref{eq:axialtorsion}) of
the torsion tensor  \cite{Shapiro:2001rz}.  Using the definitions (\ref{kands}) and
(\ref{tva}) we obtain
\bea
\label{ld2}
{L}_D &=& \frac{i}{2} e_{a}^{\ \mu} ( \bar{\psi} \gamma^{a}
\tilde{D}_{\mu}
\psi-\overline{\tilde{D}_{\mu}\psi}\gamma^{a} \psi)-m\bar{\psi}\psi
\nonumber \\
&&+\frac{3}{2}\sigma  e_{b}^{\ \mu} a_{\mu} \bar{\psi} \gamma^{5} \gamma^{b} \psi  \,.
\eea
Here $\tilde{D}_{\mu} = \partial_{\mu}
-i\sigma^{bc}g_{\rho\nu}e_{b}^{\ \nu}
\tilde{\nabla}_{\mu}e_{c}^{\ {\rho}}/4$,
 $\tilde{\nabla}_{\mu}e_{c}^{\ \rho} = \partial_{\mu} e_{c}^{\
   \rho}+    \biglb\{ \stackrel{\rho}{\scriptstyle \mu\nu}
 \bigrb\}   e_{c}^{\ \nu}$,
and $\gamma^{5} =  i\gamma^{0}\gamma^{1}\gamma^{2}\gamma^{3}$, where we use the convention $\epsilon_{0123}=\sqrt{-g}$.

Now the first line in the Dirac action (\ref{ld2}) depends only on the metric, and
in particular it is independent of $a_{\mu\nu}$, so it is trivially
invariant under the transformation (\ref{sym3}).
The second line depends on $a_{\mu\nu}$ only
through the axial piece $a_{\mu}$ of the torsion, which is given by
Eq.\ (\ref{axialpiece}), and which is also invariant under
(\ref{sym3}).  Therefore the entire action (\ref{ld2}) is invariant
under the symmetry.

We conclude that on the domain ${\cal D}_2$ we have a consistent
theory with a well posed initial value formulation, in which the
symmetry (\ref{sym3}) is a gauge symmetry.
The self-consistency of this theory for $\sigma=1$ was previously discussed by
Leclerc \cite{Leclerc:2005jr}.  From Eqs.\
(\ref{gaction1}) and (\ref{calD2def}), the complete action for the
theory is
\bea
S[e_{\ \mu}^a,\Psi] &=& \int d^4x \sqrt{-g}
\left[\frac{1}{2\hat{\kappa}}R(\{\})+  9 d_3 a_\mu a^\mu \right]
\nonumber \\&& + S_D[e_{\ \mu}^a,\psi,\bar{\psi}] + \ldots,
\eea
where ${\hat \kappa} = 1 / (2 d_1)$.
Here the ellipses denote additional terms in the standard model of
particle physics that are not coupled to the torsion; the only term
that couples to the torsion is the Dirac action for the fermions
(the ``minimal coupling'' scheme of Refs.\
\cite{Hehl:1976kj,Shapiro:2001rz}).

The linearized equation of motion for $e_{\ \mu}^a$ obtained from this
action gives equations for $h_{\mu\nu}$ and $a_{\mu\nu}$:
\bes
\label{eq:ehslinear}
\bea
\label{grlinear}
-\frac{1}{2}\Box \bar{h}_{\mu\nu} +\bar{h}_{\rho(\mu,\nu)}^{\ \ \ \ \ \ \rho}  -\frac{1}{2}\eta_{\mu\nu} \bar{h}_{\rho\lambda}^{\ \ ,\rho\lambda}&=&
\hat{\kappa}T_{(\mu\nu)},\\\label{aeq}
\Box a_{\mu\nu} +2a_{\rho[\mu,\nu]}^{\ \ \ \ \ \ \rho}
&=&\frac{1}{d_3}T_{[\mu\nu]} \,. \ \ \ \
\eea
\ees
Here $\Box=\eta^{\alpha\beta}\partial_{\alpha}\partial_{\beta}$,
$\bar{h}_{\mu\nu}\equiv h_{\mu\nu}-\eta_{\mu\nu} h_{\rho}^{\ \rho}$,
${\cal T}_{\mu\nu}$ is the non-symmetric energy-momentum tensor defined by
\be
{\cal T}_{\mu}^{\ \nu} \equiv  \frac{1}{\sqrt{-g}}
\frac{\delta S_{\rm matter} } {\delta e^{a}_{\ \nu}}e^{a}_{\ \mu},
\ee
and
$T_{\mu\nu} $ is its leading order term in a perturbative expansion (i.e. $T_{\mu\nu}$ is independent of $h_{\mu\nu}$ and $a_{\mu\nu}$).
The source terms in Eqs.\ (\ref{eq:ehslinear}) obey the conservation
laws $T_{[\mu\nu]}^{\ \ \ \, ,\nu}=0$ and $T_{(\mu\nu)}^{\ \ \ \,
  ,\nu}=0$, by virtue of the invariance of the matter action under the
transformations (\ref{diff}) and (\ref{sym3}).
By using these transformations we can impose the gauge conditions
$\bar{h}_{\mu\nu}^{\ \ \, ,\nu}=0$ and $a_{\mu\nu}^{\ \ \, ,\nu}=0$, thereby
reducing the field equations to wave equations:
\bes
\bea
\label{waveeqh}
&&  \Box \bar{h}_{\mu\nu}  = -2 {\hat \kappa} T_{(\mu\nu)},\\ \label{waveeqa}
 && \Box a_{\mu\nu}  =\frac{1}{d_3}T_{[\mu\nu]} \,.
\eea
\ees
The first of these is the usual linearized Einstein equation.

Next we examine the antisymmetric piece of the stress energy tensor
which acts as a source for $a_{\mu\nu}$ in Eq.\ (\ref{waveeqa}).
One can show \cite{Hayashi:1979qx} that this  antisymmetric piece is related to the 
divergence of the spin density tensor by
\begin{equation}\label{tdivs}
T_{[\beta\alpha]}=  \sigma b^a_{\  \alpha} b^b_{\ \beta} [\sigma_{ab \ ,\mu}^{\ \ \mu}]_{b_c^{\ \nu}}\,,
\end{equation}
where the subscript ${b_c^{\ \nu}}$ indicates evaluation at the background values
of the tetrad. Recall that the spin density is defined by 
$$
\sigma_{ab}^{\ \ \mu} \equiv \frac{1}{\sqrt{g}} \frac{ \delta S_{\rm    matter}[e^f_{\  \alpha},\omega_\nu^{\ cd},\Psi]}
 {\delta \omega_\mu^{\ ab}},
$$
where  in this definition the matter action is considered to 
be a functional of the {\em independent} variables 
$e^f_{\  \alpha}$ , $\omega_\nu^{\ cd}$ and $\Psi$.  The matter action can be brought to this  desired 
form by  substituting $D_{\mu} = \partial_{\mu}-\frac{i}{4}\sigma^{bc}g_{\rho\mu} \omega ^{\rho}_{\ bc}$ 
into  the expression for $L_D$ in  Eq. (\ref{ld1}). 
Relation (\ref{tetconn}) guarantees that this expression for $D_{\mu}$  equals to our 
original expression in  Eq. (\ref{Dmudef}). 

Equations (\ref{waveeqa}) and (\ref{tdivs}) imply that $a_{\mu\nu}$ is sourced only by the intrinsic spin density of matter.
As we have discussed, integrating Eq.\ (\ref{waveeqa}) for a
macroscopic object for which the spins of the elementary particles are
not aligned over a macroscopic scale gives a
negligible $a_{\mu\nu}$ \cite{Hayashi:1979qx}, and consequently
the predictions of the linearized theory coincide with those of GR.
Thus there is no extra torsion-related signal predicted for GPB for
this theory.

The lack of an experimental signature of torsion may seem strange,
since the torsion tensor is non-vanishing even in the limit where one can
 neglect intrinsic spin. 
As discussed in the introduction, the explanation
is that the torsion is not an independent dynamical degree of freedom.  More
specifically, to linear order, macroscopic bodies give rise to a metric perturbation $h_{\mu\nu}$
in the same way as in GR, and then the torsion is simply defined to be
\be
S_{\mu\nu}^{\ \ \lambda} = \frac{\sigma}{2}  h^{\lambda}_{\
    [\nu,\mu]}
\label{phantomtorsion}
\ee
[c.f.\ the first term in Eq.\ (\ref{torsionf})].
This definition has no dynamical consequence, since
only the axial piece (\ref{eq:axialtorsion}) of torsion couples to
matter, and the expression (\ref{phantomtorsion}) has no axial piece.

\section{Conclusions}\label{conc}

Preliminary results from Gravity Probe B will be announced in April 2007.
The primary scientific goals of the experiment are to verify the predictions of
general relativity for geodetic precession and for dragging of inertial frames \cite{GPB}.
However the mission is also potentially useful as a probe of modifications of general relativity.

One class of theories of gravity that GPB could potentially usefully constrain are theories
involving a dynamical torsion tensor.  Mao et.\ al.\ suggested a particular class of torsion
theories that they argued would predict a measurable torsion signal for GPB \cite{mtgc}.
We have shown that this particular class of theories is internally consistent in only a
small region of its parameter space, and in that consistent region does not predict any signal for GPB.
There may exist other torsion theories which could be usefully
constrained by GPB.  It would be interesting to find such theories.

\acknowledgments

This research was sponsored in part by NSF grant PHY-0457200.

\appendix




\section{Ghosts in the Linearized theory}
\label{sec:ghost1}

In this Appendix we show that the Hamiltonian of the complete linearized EHS theory
(\ref{linaction}) is unbounded below whenever the corresponding
Hamiltonian of the antisymmetric term (\ref{la}) is unbounded from
below. This allows us to deduce the existence of ghosts 
in the complete theory from their existence in the reduced theory (\ref{la}).

It is sufficient to show this property for the case of finite
dimensional, quadratic dynamical systems.  We consider a system with
$N$ degrees of freedom whose Lagrangian and Hamiltonian are given  by
\begin{eqnarray}
&&L=\frac{1}{2} \dot{q}_m \dot{q}_n K_{mn} -V(q_m)\,,\\\label{fullh}
&&H=\frac{1}{2} {p}_m {p}_n {(K^{-1})}_{mn}+V(q_m)\,.
\end{eqnarray}
Here $q_m$ denotes the generalized coordinates and $p_m= K_{mn}
\dot{q}_n$ denotes the conjugate momenta, where the indices $m,n$ run
from 1 to N, and $K_{mn}=K_{nm}$.  Recall that the antisymmetric
action $S_A$ was obtained by ignoring some of the dynamical variables
of the linearized EHS theory.
In a discrete theory this corresponds to writing a reduced Lagrangian $L_r$ that depends  only on some
of generalized coordinates $q_i$, where  $i$ runs from 1 to $M$, $M<N$.
The reduced theory has the following Lagrangian and Hamiltonian
\begin{eqnarray}
&&L_r=\frac{1}{2} \dot{q}_i \dot{q}_j k_{ij} -U(q_i)\,,\\\label{hr}
&&H_r=\frac{1}{2} \tilde{p}_i \tilde{p}_j {(k^{-1})}_{ij}+U(q_i)\,.
\end{eqnarray}
Here  $\tilde{p}_i= k_{ij} \dot{q}_j$, where $k_{ij}=K_{ij}$ for
$i,j=1... M$ and
\be
U(q_1,....q_M)=V(q_1....q_M,0 ..0),
\label{eq:Udef}
\ee
i.e., the potential $U(q_i)$
is obtained from $V(q_m)$ by setting $q_m=0$ for $m=M+1...N$.

Now suppose that the potential term $U(q_i)$  
of the reduced Hamiltonian $H_r$ is unbounded from below.  
It follows immediately from the
definition (\ref{eq:Udef}) that $V(q_m)$ is also unbounded from below,
and so the complete Hamiltonian (\ref{fullh}) is unbounded from below.

Suppose next that the kinetic term of the reduced Hamiltonian $H_r$ 
is unbounded from below.  By virtue of Eq.\ (\ref{hr})  this means that
 at least one of the eigenvalues  of $(k^{-1})_{ij}$ must be negative.
Denoting the eigenvalues of the matrix $k_{ij}$ by $\lambda_{(i)}$, $1
\le i \le M$, we find that there exists an eigenvalue $l$ for which
 $\lambda_{(l)}^{-1}<0$.
This implies that there exists an M dimensional eigenvector
$w^{(l)}_{i}$ for which $w^{(l)}_{i} w^{(l)}_{j}
k_{ij}=\lambda_{(l)}<0$. We now construct
an N dimensional vector defined by $\eta_m=(w^{(l)}_1...w^{(l)}_M,0...0)$. By definition, this vector satisfies $\eta_m \eta_n K_{mn}=\lambda_{(l)}<0$, implying that
$K_{mn}$ has a negative eigenvalue.
This means that the complete Hamiltonian (\ref{fullh}) 
is unbounded from below.

\section{Alternative form of gravitational action}
\label{sec:identity}

In this Appendix we derive identity
\bea
\label{identity}
\int \sqrt{-g} R(\{\}) d^4x &=&\int \sqrt{-g}
\bigg[ - \frac{8}{3} t_{\mu\nu\lambda} t^{\mu\nu\lambda} \nonumber \\
&&+ \frac{8}{3} v_\nu v^\nu -6 a_\nu a^\nu \bigg],
\eea
where $R(\{\})$ is the Ricci scalar of the Levi-Civita connection, and
$t_{\mu\nu\lambda}$, $v^\mu$ and $a^\mu$ are the irreducible pieces
(\ref{tva}) of the torsion tensor with the factor of $\sigma$ removed.
Combining this identity with the formula (\ref{gaction}) for the
gravitational action of the EHS theory yields the alternative form
(\ref{gaction1}) of that action.

The idea is to introduce a new torsion tensor
\begin{equation}
\label{torsionbar}
{\bar S}_{\mu\nu}^{\ \ \lambda} \equiv \frac{1}{2}e_{a}^{\ \lambda} (e^{a}_{\ \nu,\mu}- e^{a}_{\ \mu,\nu} )\,.
\end{equation}
This is just the torsion tensor (\ref{torsion}) of the EHS theory but
specialized to $\sigma=1$, i.e., it is the torsion tensor of the
Hayashi-Shirafuji teleparallel theory\cite{Hayashi:1979qx}.
From Eqs.\ (\ref{torsion}) and (\ref{tva}) it is related to the fields $t_{\mu\nu\lambda}$,
$v^\mu$ and $a^\nu$ by
\begin{equation}\label{stva}
\bar{S}_{\nu\mu\lambda}=\frac{2}{3}(t_{\lambda\mu\nu}-t_{\lambda\nu\mu})+
\frac{1}{3}(g_{\lambda\mu}v_\nu -g_{\lambda\nu} v_{\mu})+\epsilon_{\lambda\mu\nu\rho}a^\rho\,.
\end{equation}
We denote the corresponding metric compatible  connection by $\bar{\Gamma}^\alpha_{\ \beta\gamma}$
and the corresponding Riemann tensor by ${\bar R}^\mu_{\ \nu\lambda\sigma}$.  This Riemann
tensor vanishes identically by virtue of the definition
(\ref{torsionbar}), as we discussed in the introduction.
The Ricci scalar $\bar{R}\equiv g^{\beta\delta}\bar{R}^\alpha_{\
  \beta\alpha\delta}$ also vanishes, which implies
\begin{equation}\label{barricci}
\int \sqrt{-g} \bar{R} d^4x =0\,.
\end{equation}

We now substitute into Eq.\ (\ref{barricci}) the formula
\begin{equation}\label{barriemann}
\bar{R}^\alpha_{\ \beta\gamma\delta}=\bar{\Gamma}_{\ \delta\beta,\gamma}^\alpha-
\bar{\Gamma}_{\ \gamma\beta,\delta}^\alpha+
\bar{\Gamma}_{\ \gamma\mu}^\alpha \bar{\Gamma}_{\ \delta\beta}^\mu-
\bar{\Gamma}_{\ \delta\mu}^\alpha \bar{\Gamma}_{\ \gamma\beta}^\mu\,,
\end{equation}
together with barred versions of Eqs.\ (\ref{kdef}) and (\ref{kands}).
This gives
\begin{equation}\label{i2}
\int \sqrt{-g} [R(\{\})-2 {\tilde \nabla}_\alpha \bar{K}^{\beta\ \alpha}_{\ \beta}-\bar{K}_{\ \alpha\mu}^{\alpha}\bar{K}_{\ \delta}^{\delta\ \mu}+
\bar{K}_{\ \mu}^{\beta\ \alpha} \bar{K}_{\ \mu}^{\alpha\ \beta}]=0
\end{equation}
Here ${\tilde \nabla}_\alpha$ denotes the
the Levi-Civita derivative operator.
Discarding the total derivative term and using the decomposition (\ref{stva}) together with the barred
version of Eq.\ (\ref{kands}) now yields the desired identity
(\ref{identity}).

\newcommand{\apjl}{Astrophys. J. Lett.}
\newcommand{\aap}{Astron. and Astrophys.}
\newcommand{\cmp}{Commun. Math. Phys.}
\newcommand{\grg}{Gen. Rel. Grav.}
\newcommand{\lr}{Living Reviews in Relativity}
\newcommand{\mnras}{Mon. Not. Roy. Astr. Soc.}
\newcommand{\pr}{Phys. Rev.}
\newcommand{\prsl}{Proc. R. Soc. Lond. A}
\newcommand{\ptrsl}{Phil. Trans. Roy. Soc. London}


\end{document}